\documentclass[12pt]{article}
\usepackage{latexsym}
\usepackage{citesort,rotate}
\input epsf

\newcommand{\sect}[1]{\setcounter{equation}{0}\section{#1}}

\begin{document}

\title{Quantum global structure of de Sitter space}

\author{{\sc Raphael Bousso}\thanks{\it
       bousso@stanford.edu}
       \\[1 ex] {\it Department of Physics}
       \\ {\it Stanford University}
       \\ {\it Stanford, California 94305-4060}
       }

\date{SU-ITP-99-7~~~~25 February 1999~~~~hep-th/9902183}

\maketitle

\begin{abstract}

I study the global structure of de~Sitter space in the semi-classical
and one-loop approximations to quantum gravity. The creation and
evaporation of neutral black holes causes the fragmentation of
de~Sitter space into disconnected daughter universes. If the black
holes are stabilized by a charge, I find that the decay leads to a
necklace of de~Sitter universes (`beads') joined by near-extremal
black hole throats. For sufficient charge, more and more beads keep
forming on the necklace, so that an unbounded number of universes will
be produced. In any case, future infinity will not be connected. This
may have implications for a holographic description of quantum gravity
in de~Sitter space.

\end{abstract}

\pagebreak

\sect{Introduction}

\subsection{Why study de~Sitter space?}
\label{sec-why}

In an explicit realization of the holographic
principle~\cite{SteTho94,Sus95,SusWit98}, it has recently been argued
that type IIB string theory in anti-de~Sitter space is dual to a
conformal field theory on its boundary~\cite{Mal97,Wit98a,Wit98b}.
Anti-de~Sitter space is the maximally symmetric solution of the vacuum
Einstein equations with a negative cosmological constant $\Lambda<0$.
One would expect different implementations of the holographic
principle to apply to Minkowski ($\Lambda=0$) and de~Sitter space
($\Lambda>0$), but explicit prescriptions have yet to be found.

\begin{figure}[htb!]
  \hspace{.25\textwidth} \vbox{\epsfxsize=.5\textwidth
    \rotate[r]{\epsfbox{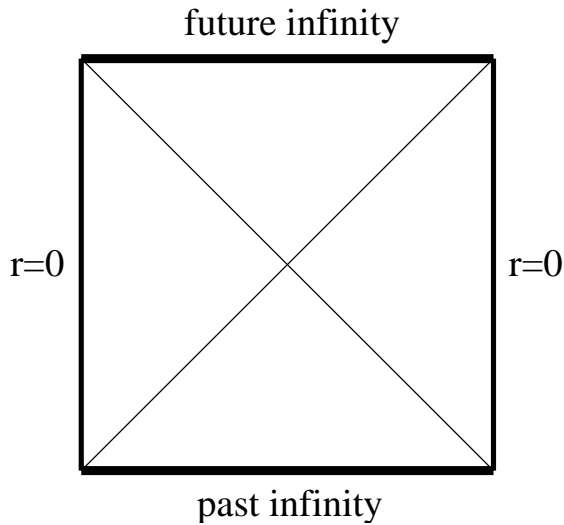}}}
\caption%
{\small\sl Penrose diagram of classical de~Sitter space. The vertical
  lines are the origins of polar coordinates on opposite poles of the
  spatial three-spheres. The diagonal lines are the event horizons of
  observers on these poles.}
\label{fig-cp-ds}
\end{figure}

An important open question in the latter cases is the location of the
surfaces onto which the gravitational theory should be projected.
de~Sitter space contains no boundaries at spatial infinity, but it has
spacelike boundaries at past and future infinity (see
Fig.~\ref{fig-cp-ds}).  Witten has suggested~\cite{Wit98SB} that these
boundaries may provide the holographic surfaces for de~Sitter space.

It is therefore crucial to understand whether the simple global
structure of de~Sitter space persists when quantum effects are
included. Because de~Sitter space is at a non-zero temperature, black
holes will necessarily form through non-perturbative quantum
fluctuations of the metric~\cite{GinPer83,BouHaw95}. Therefore the
future endpoints of some geodesics fail to lie on future infinity;
instead, some will end on black hole singularities. Moreover, the
evaporation of some of these black holes will cause the spacelike
surfaces to fragment into large, disconnected
pieces~\cite{Bou98}. This means that future infinity will not even be
connected. Since the daughter universes are locally de~Sitter, one
might expect the process to repeat there. This would cause the
proliferation into an infinite number of separate de~Sitter universes,
and thus the complete fragmentation of future infinity.

The proliferation effect exploits a local instability of the
maximum-size neutral black holes created in de~Sitter space.  The
present paper will expose a similar instability in maximal black holes
carrying magnetic charge. They also nucleate spontaneously in
de~Sitter space.  Similar to the uncharged case, the instability leads
to a necklace of de Sitter regions connected through black hole
throats. Since the black holes will be stabilized near the extremal
limit, however, they will not evaporate completely. In the charged
case, therefore, the necklace will not break up into separate
de~Sitter spaces; instead, the daughter universes will remain
connected through near-extremal black holes. If the process repeats in
them, a network of such necklaces will emerge. Moreover, if the black
holes are sufficiently charged, they will not evaporate at all. In
this case the near-black hole regions will remain forever classically
unstable. They will produce an unbounded number of black hole
interiors and de~Sitter `beads.' The inclusion of charge thus makes
the semi-classical global structure of de~Sitter space even richer.

In order to avoid cluttering the equations with factors of $(D-i)$,
only four-dimensional de~Sitter space will be considered here. Maximal
black hole solutions exist for all de~Sitter spaces with spacetime
dimension $D \geq 4$, and should lead to the same instabilities and
global structure.

de~Sitter space is an important solution of Einstein's equations, and
the semiclassical investigation of its quantum properties is useful
both in its own right and as a potential guide in the search for a
complete quantum description. An additional motivation was emphasized
in Refs.~\cite{BouHaw95,BouHaw96,Bou96,Bou96b,Bou98}: The universe
presumably underwent a primordial inflationary era, during which it
behaved much like de~Sitter space. If inflation was sufficiently long
(and it is, in fact, generically eternal~\cite{Lin86a}), such quantum
effects will determine the global structure of the universe. Recent
supernova measurements~\cite{Per98,FilRie98} indicate, moreover, that
the universe has once again entered a vacuum-dominated era, albeit
with a much smaller (effective) cosmological constant. If this vacuum
energy is stable, the results found here will be of relevance also for
the distant future of the universe.

\subsection{Proliferation}
\label{sec-conditions}

The fragmentation of de~Sitter space was found in Ref.~\cite{Bou98}
for four-dimensional Einstein gravity with a positive cosmological
constant and no Maxwell field. Related effects were later investigated
for other theories in Refs.~\cite{NojOdi98c,NojOdi98d,BytNoj98}. It
involves three steps, which can be qualitatively described as follows
(see Fig.~\ref{fig-steps}):
\begin{figure}[htb!]
  \hspace{.2\textwidth} \vbox{\epsfxsize=.6\textwidth
  \epsfbox{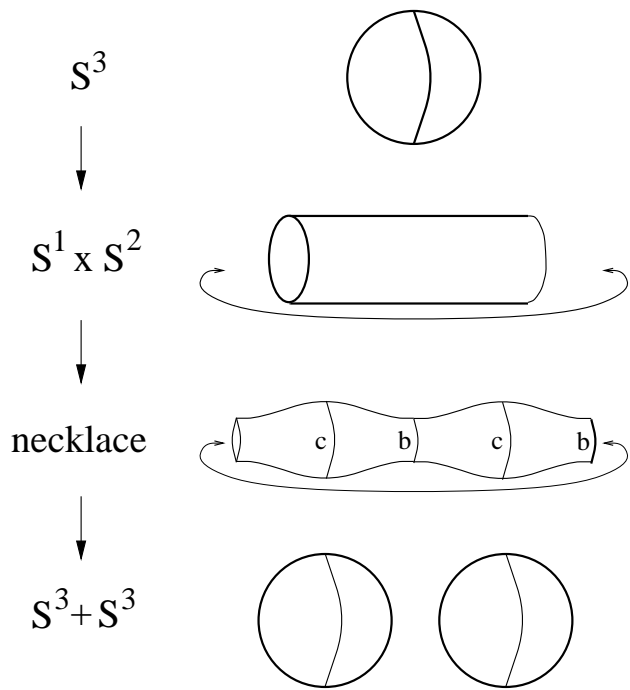}}
\caption%
{\small\sl The three-step process by which de~Sitter space
  proliferates into disconnected daughter universes. On a de~Sitter
  background with $S^3$ spacelike sections, a neutral Nariai solution
  ($S^1 \times S^2$) nucleates (the arrows indicate that opposite ends
  should be identified, to form the $S^1$). By a classical
  instability, the degenerate Nariai geometry decays into a necklace
  of $n$ black holes and $n$ de~Sitter regions (`beads'). Here $n=2$,
  and the spatial section was chosen so that it penetrates neither the
  black hole interiors nor the de Sitter regions; this means that the
  minimal two-spheres correspond to the black hole horizons (b), and
  the maximal two-spheres correspond to the cosmological horizons (c).
  When the black holes evaporate, the beads disconnect, and $n$
  separate de~Sitter universes remain. The corresponding Penrose
  diagram is shown in Fig.~\ref{fig-cp-prolneut}.}
\label{fig-steps}
\end{figure}

\begin{enumerate}

\item{In de~Sitter space, a non-perturbative, instanton-mediated
fluctuation of the gravitational field changes the spatial topology
from $S^3$ to $S^1 \times S^2$.  This transition occurs at a rate of
$e^{-\pi/\Lambda}$ and leads to the maximal Schwarzschild-de~Sitter
solution (a.k.a.\ Nariai), in which the $S^2$ radius is independent of
the $S^1$ coordinate.}

\item{Quantum fluctuations will break the degeneracy of the two-sphere
radius. The Nariai geometry is unstable under such
perturbations. Regions along the $S^1$ in which the two-spheres are
smaller than the degenerate size will collapse and form interiors of
black holes. Where the two-spheres are greater, they will grow into
exponentially large, asymptotically de~Sitter regions. The geometry
can now be thought of as a necklace, with large de~Sitter beads
connected through black hole interiors. The number of beads depends on
the number of oscillations of the initial perturbation.}

\item{The inclusion of quantum radiation at the one-loop level shows
that the black holes radiate and decrease in size. For neutral black
holes, there is no known obstruction to their complete
evaporation. This pinches the necklace, leaving only the disconnected
beads, i.e., separate de~Sitter universes.}

\end{enumerate}

\begin{figure}[htb!]
  \hspace{.05\textwidth} \vbox{\epsfxsize=.9\textwidth
  \epsfbox{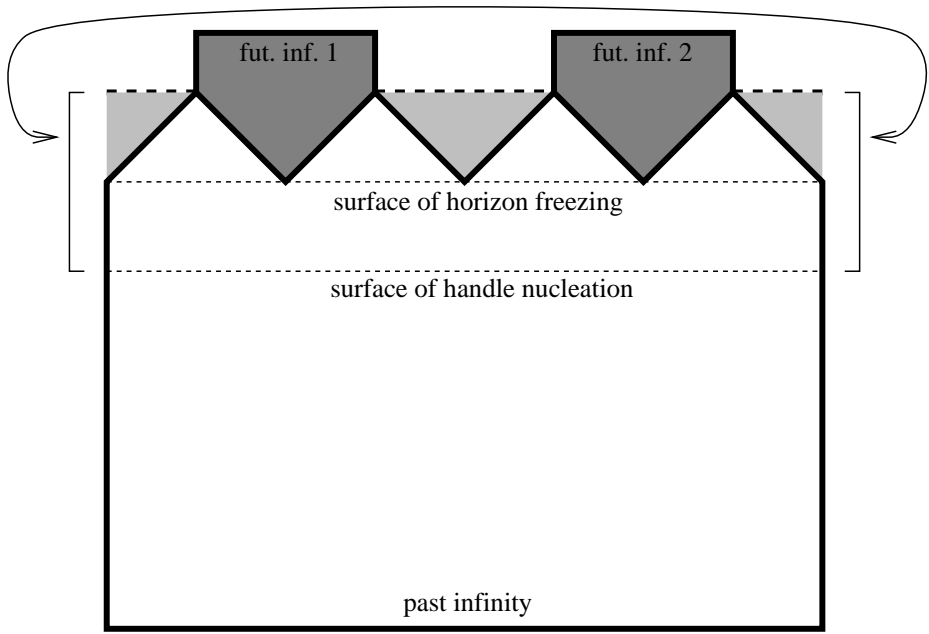}}
\caption%
{\small\sl Penrose diagram for the proliferation process depicted in
  Fig.~\ref{fig-steps}. Singularities are indicated by dashed lines.
  In the region marked by the square brackets the spatial topology is
  $S^1 \times S^2$, and opposite ends should be identified.
  Perturbations of the two-sphere radius first oscillate, then freeze
  out, seeding black hole interiors (light gray) and de~Sitter regions
  (dark grey). After the black holes evaporate, two separate de~Sitter
  universes remain.}
\label{fig-cp-prolneut}
\end{figure}

The causal structure that results from this process is shown in
Fig.~\ref{fig-cp-prolneut} for $n=2$.

\subsection{Outline and Summary}
\label{sec-outline}

I will assume that there are no magnetically charged particles, or at
least none of smaller mass than the black holes. This means that
magnetic black holes cannot radiate away their charge. If magnetic
flux was threaded around the $S^1$, the black holes could no longer
evaporate completely, because they would have to retain at least the
minimum mass necessary to support their charge. In other words, the
necklace could not be pinched. The large de~Sitter universes would
have to remain connected through near-extremal black hole throats.

My aim is to show that ordinary de~Sitter space decays into this type
of stable `necklace' configuration.  This will be an additional,
`charged' decay mode complementing the proliferation effect that is
mediated by neutral black holes.  For this purpose, I will investigate
the stability of a class of charged black holes in de~Sitter space.
The two-parameter family of Reissner-Nordstr\"om-de~Sitter solutions
will be reviewed in Sec.~\ref{sec-rnds}.  For a black hole to nucleate
spontaneously (step 1), its geometry must contain a smooth Euclidean
sector (see Ref.~\cite{BouHaw98} for a discussion and possible
exceptions from this rule). This restricts to a solution subspace
comprising three one-parameter families, the `cold', `lukewarm', and
`Charged Nariai' black holes~\cite{ManRos95}.

Multiple black hole interiors and de~Sitter regions (step 2) can only
form if the spatial topology contains an $S^2$ factor {\em of constant
  radius}. In the cold and lukewarm solutions, the two-sphere size
will vary. Its single minimum will be distinctly smaller than its
single maximum. When one starts with classical solutions of this type,
small quantum fluctuations can have no influence on the number of
oscillations about the degenerate two-sphere size. The necklace will
contain only one bead, whose opposite ends will be connected by a
single black hole interior. The production of such black holes is
certainly interesting; it affects the quantum structure of de~Sitter
space in the sense that not all world-lines will end on future
infinity. But it is not new.  Here, therefore, the focus will be on
the Charged Nariai solutions, for which the two-sphere size is
classically constant, and the number of oscillations is determined
only by quantum fluctuations. Their metric and causal structure are
presented in Sec.~\ref{sec-cn} (In the neutral case, both step 1 and
step 2 restrict the Schwarzschild-de~Sitter solutions to the one with
the largest black hole size, the neutral Nariai solution.)

It is important to distinguish between the classical and the quantum
instability of the Charged Nariai solutions. Classically, they are
perturbatively unstable to the transition into a
Reissner-Nordstr\"om-de~Sitter solution representing a black hole of
nearly maximal mass (Sec.~\ref{sec-prelim-cl}). This solution, in
turn, may decay into a lower mass black hole by emitting quantum
radiation. In Sec.~\ref{sec-prelim-th} I will argue, on thermodynamic
grounds only, that maximal black holes of sufficiently large, yet
sub-extremal charge will not possess the quantum instability.

Classically, the constraint equations allow only fluctuations that
lead to the formation of a single black hole. In a one-loop effective
model that includes quantum radiation (Sec.~\ref{sec-model}), however,
this restriction is lifted.  In Sec.~\ref{sec-solutions}, I will show
that higher-mode perturbations of a Charged Nariai solution first
oscillate, then freeze out and form multiple black holes and de~Sitter
regions. The model includes the back-reaction to the quantum radiation.
It predicts that the horizons do not, at first, behave according to
expectations developed in Sec.~\ref{sec-prelim-th}. At later times,
however, they do follow the thermodynamic behavior. In fact, the
one-loop model turns out to reproduce exactly the value of the
critical charge, beyond which maximal black holes no longer evaporate.

\begin{figure}[htb!]
  \hspace{.05\textwidth} \vbox{\epsfxsize=.9\textwidth
    \epsfbox{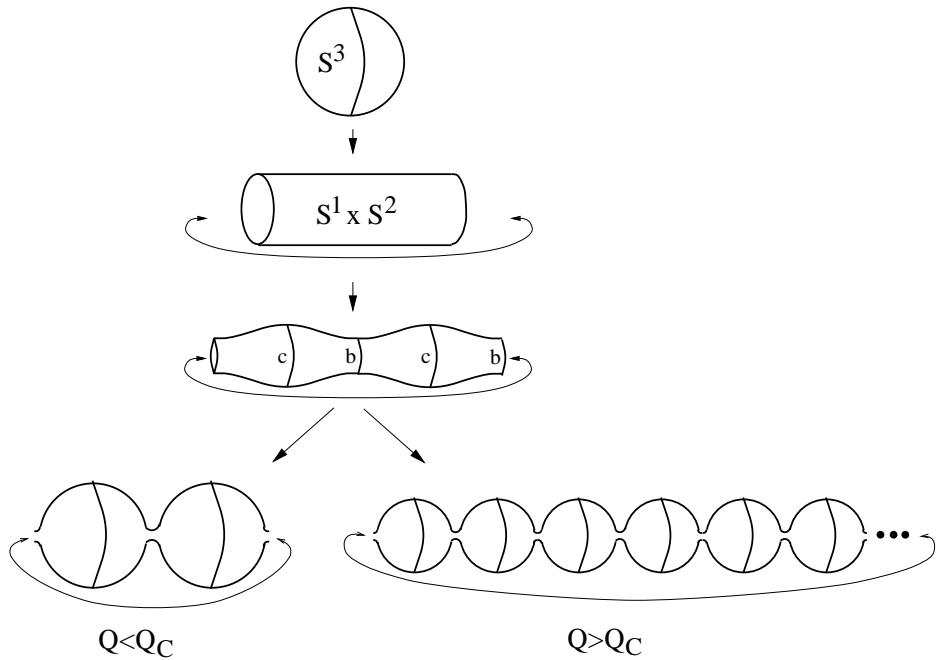}}
\caption%
{\small\sl Charged Nariai black holes nucleate semiclassically in
  de~Sitter space. Their spacelike sections are $S^1 \times S^2$.  The
  one-sphere expands exponentially, and the two-sphere has a constant
  radius. The magnetic field loops around the one-sphere and prevents
  it from being pinched. Quantum fluctuations lead to the formation of
  $n$ black holes and $n$ de~Sitter regions (here $n=2$).  If the
  charge is small, the black holes evaporate until they are nearly
  extremal (see Fig.~\ref{fig-cp-prolsub} for a Penrose diagram). For
  supercritical charge, the black holes will grow, approaching the
  radius of the Charged Nariai solution. The regions between a black
  hole and a cosmological horizon remain nearly degenerate. Small
  perturbations can produce more black hole interiors and de~Sitter
  regions there. Iteratively, an infinite number of beads will
  form. The corresponding Penrose diagram is shown in
  Fig.~\ref{fig-cp-prolsup}.}
\label{fig-structure}
\end{figure}
\begin{figure}[htb!]
  \hspace{.05\textwidth} \vbox{\epsfxsize=.9\textwidth
  \epsfbox{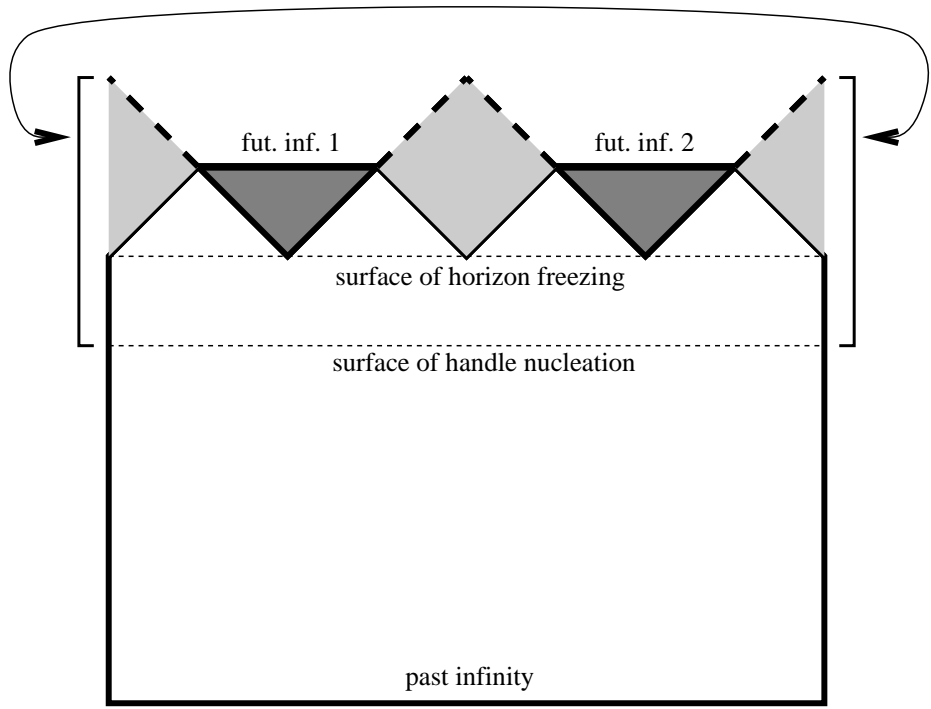}}
\caption%
{\small\sl Penrose diagram for the decay of de~Sitter space into $n$
  de~Sitter regions (dark grey) separated by $n$ lukewarm black holes
  (light gray); here $n=2$. Singularities are indicated by dashed
  lines. The upper part of the diagram is a sequence of $n$
  Reissner-Nordstr\"om-de~Sitter diagrams (Fig.~\ref{fig-cp-rnds-tr}).
  Future infinity fragments, but in contrast to the neutral case
  (Fig.~\ref{fig-cp-prolneut}), space remains topologically
  connected. This diagram corresponds to the subcritical branch of
  Fig.~\ref{fig-structure}.}
\label{fig-cp-prolsub}
\end{figure}

The different evolution of sub- and supercritically charged black
holes leads to a drastic difference in their global structure.  Black
holes of subcritical charge simply radiate mass away until they are
nearly extremal. The final picture is similar to neutral
proliferation, except that the large de~Sitter beads will still be
connected through small black holes; see Fig.~\ref{fig-structure}
(left branch).  The corresponding Penrose diagram is shown in
Fig.~\ref{fig-cp-prolsub}. Like in the neutral case, the number of
final de~Sitter universes is determined by the mode number of the
dominant perturbation when the Charged Nariai solution nucleates.

\begin{figure}[htb!]
  \hspace{.0\textwidth} \vbox{\epsfxsize=1.0\textwidth
  \epsfbox{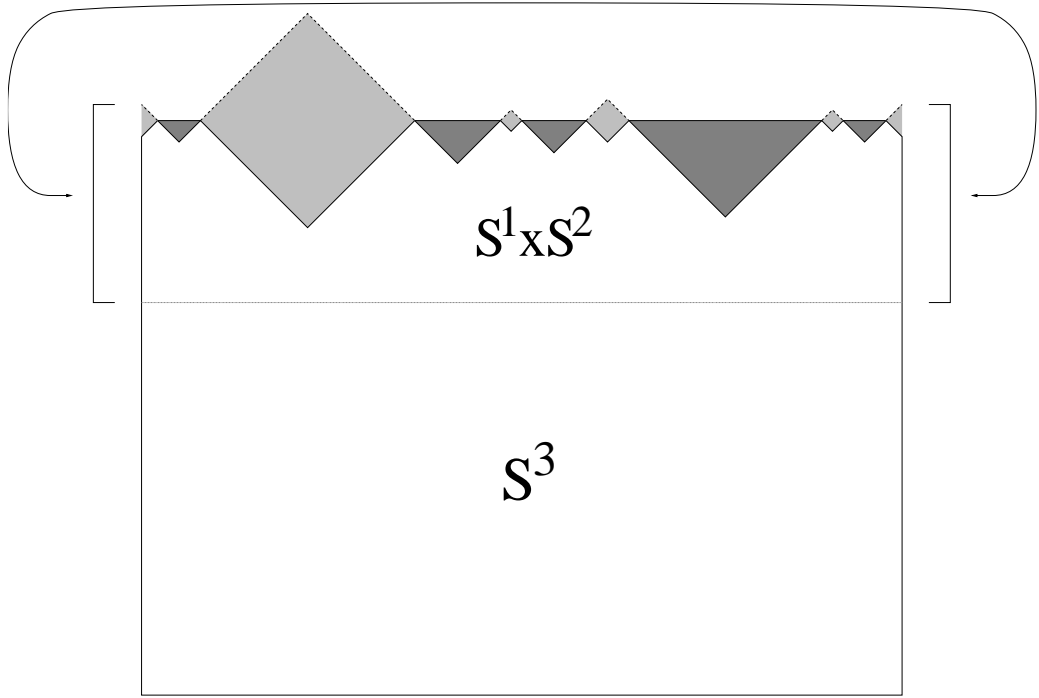}}
\caption%
{\small\sl Nucleation of a supercritically charged Nariai
solution. Black holes form but do not evaporate. Between a black hole
and a cosmological horizon, the $S^2$ size remains nearly constant,
and small fluctuations can produce more black holes (light grey) and
de~Sitter regions (dark grey). In this way, an infinite number of
beads form on the $S^1$. With better resolution, one would see more
and more diamonds and triangles towards the top. This Penrose diagram
corresponds to the supercritical branch of Fig.~\ref{fig-structure}.}
\label{fig-cp-prolsup}
\end{figure}

For Charged Nariai solutions of supercritical charge, an amusing new
effect occurs.  They will still form large de~Sitter beads as well as
black hole interiors with singularities. But the black holes do not
evaporate. Instead, their horizon size asymptotically approaches the
Charged Nariai value.  Therefore there will always be regions where
the two-sphere size is nearly constant, namely the regions between
each black hole horizon and the cosmological horizon `surrounding' it.
Random quantum fluctuations of the two-sphere size in such regions
will include perturbations which have a maximum on the black hole side
and a minimum on the side of the cosmological horizon. If the
fluctuation is strong enough, the two-spheres will be larger than the
Charged Nariai radius at the maximum, and smaller at the minimum. As
it crosses the apparent cosmological horizon, the minimum will freeze
out and collapse, forming a new black hole interior. Similarly, the
maximum will freeze out as it crosses the apparent black hole horizon;
it then grows exponentially, seeding a new de~Sitter region.
Effectively, a new bead will have been inserted into the necklace. The
process continues iteratively, so that infinitely many black holes and
de~Sitter regions will form around the $S^1$. The result is an
unbounded number of de Sitter universes, topologically (but not
causally) connected by charged wormholes (see
Fig.~\ref{fig-structure}, right branch).  The corresponding Penrose
diagram is a fractal; it is sketched in Fig.~\ref{fig-cp-prolsup}.

One might have expected the proliferation of de~Sitter space via
charged black holes to offer nothing new beyond the additional rule
that the daughter universes never quite disconnect. The above argument
shows, however, that in fact a genuinely new phenomenon arises.  When
a neutral or subcritically charged Nariai solution nucleates
semiclassically, at most a finite number of de~Sitter universes can
develop. Infinite proliferation occurs only if the process repeats
iteratively inside the daughter universes. This is plausible, but not
entirely obvious (see Sec.~\ref{sec-discussion}). But in the
supercritical case, an infinite number of de~Sitter beads forms on a
single $S^1$ necklace.

In Sec.~\ref{sec-discussion}, I will question the global viewpoint
that pervades this paper. I will argue that at least some of my
results are independent of its validity, and examine how its
abandonment may affect the location of holographic surfaces.

\sect{Reissner-Nordstr\"om-de~Sitter black holes}
  \label{sec-classical}

\subsection{General Lorentzian solution}
  \label{sec-rnds}

The four-dimensional Lorentzian Einstein-Hilbert action with a
cosmological constant, $\Lambda$, and a Maxwell field, $F_{\mu\nu}$,
is given by:
\begin{equation}
S = \frac{1}{16 \pi} \int d^4\!x\, (-g^{{\rm IV}})^{1/2} \left[
 R^{{\rm IV}} - 2 \Lambda - F_{\mu\nu} F^{\mu\nu}
 - \frac{1}{2} \sum_{i=1}^{N}
 (\nabla^{{\rm IV}} f_i)^2 \right],
\label{eq-action-4D}
\end{equation}
where $R^{{\rm IV}}$ and $g^{{\rm IV}}$ are the four-dimensional Ricci
scalar and metric determinant. The scalar fields $f_i$ will be
needed later to carry the quantum radiation; classically, they can be
set to zero.

The charged, static, spherically symmetric solutions of the vacuum
Einstein equations with a cosmological constant $\Lambda$ are given by
the Reissner-Nordstr\"om-de~Sitter metric
\begin{equation}
ds^2 = - V(r) dt^2 + V(r)^{-1} dr^2 + r^2 d\Omega^2,
\label{eq-rnds}
\end{equation}
where
\begin{equation}
V(r) = 1 - \frac{2\mu}{r} + \frac{Q^2}{r^2} - \frac{\Lambda}{3} r^2;
\end{equation}
$d\Omega^2$ is the metric on a unit two-sphere, $Q$ is the charge, and
$\mu$ is a mass parameter. The black holes can be either magnetically
charged,
\begin{equation}
F = Q \sin \theta \, d\theta \wedge d\phi
\label{eq-max-magnetic}
\end{equation}
(whence $F^2 = 2 Q^2/r^4$), or electrically charged,
\begin{equation}
F = \frac{Q}{r^2} dt \wedge dr
\label{eq-max-electric}
\end{equation}
(whence $F^2 = -2 Q^2/r^4$).
Electrically charged black holes can lose their charge by emitting
electrically charged particles. But the aim here is to stabilize the
black holes through their charge. Therefore only magnetically charged
black holes will be considered below.
\begin{figure}[htb!]
  \hspace{.15\textwidth} \vbox{\epsfxsize=.7\textwidth
  \epsfbox{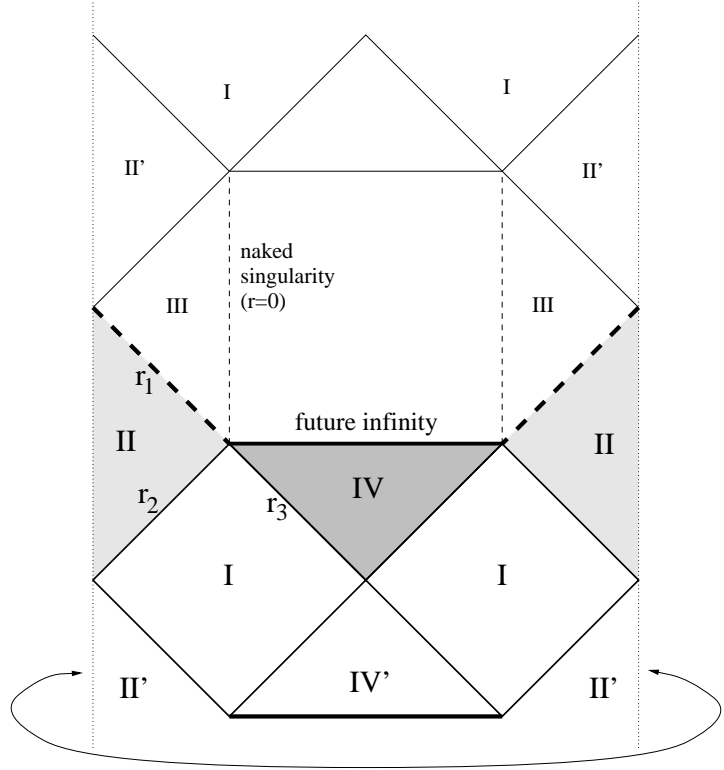}}
\caption%
{\small\sl Penrose diagram of a generic Reissner-Nordstr\"om-de~Sitter
  solution. Every point represents a two-sphere of radius $r$. Thus
  the spatial topology is $S^1 \times S^2$. The spacetime contains an
  asymptotically de~Sitter region (dark grey, IV), bounded by the
  cosmological horizon at $r_3$. The black hole interior (light
  grey) is bounded by an `outer' horizon at $r_2$.  Back-reaction
  truncates the diagram at the `inner' horizon, $r_1$.  The
  fully extended classical diagram is indicated by thin lines.}
\label{fig-cp-rnds-tr}
\end{figure}

Within appropriate ranges for the black hole mass and charge, $V$ has
three positive roots, $r_1 < r_2 < r_3 $, which may be interpreted as
the inner and outer black hole horizons and the cosmological horizon.
Figure~\ref{fig-cp-rnds-tr} shows the causal structure of a generic
Reissner-Nordstr\"om-de~Sitter solution. In the fully extended
classical solution the black hole is traversable. It contains a naked
singularity at $r=0$ in region III, beyond the inner horizon ($r_1$).
Consider an observer in region I, between the black hole and the
cosmological horizon, and located where the cosmological acceleration
and the black hole attraction balance exactly. This observer exists
for an infinite proper time, during which he always sees radiation
coming from the cosmological horizon and falling into the black hole.
(At late times, he also sees an equal amount radiation coming out of
the black hole, which explains why the back-reaction remains small in
region I.)  An observer in region II, however, will see only the
radiation emitted by the cosmological horizon, and it will take him
only a finite proper time to cross this infinite amount of radiation.
This means that back-reaction to the quantum radiation effectively
places a null singularity on the inner horizon, and truncates the
spacetime there.

\begin{figure}[htb!]
  \hspace{.12\textwidth} \vbox{\epsfxsize=.7\textwidth
  \epsfbox{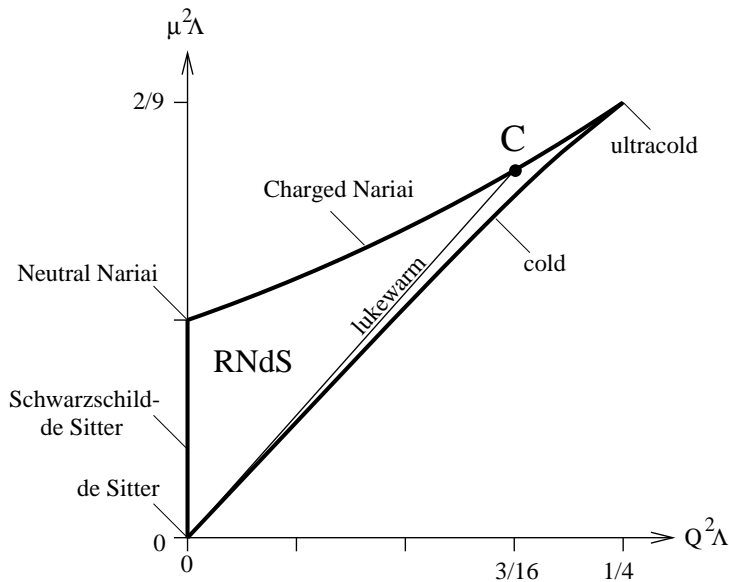}}
\caption%
{\small\sl Reissner-Nordstr\"om-de~Sitter solutions exist for all
points on or within the thick line. The plot is of the dimensionless
quantities $\mu^2 \Lambda$ vs.\ $Q^2 \Lambda$. The lukewarm solutions
lie on the dashed line, $\mu^2 = Q^2$. They are stable endpoints of
black hole evaporation. They meet the Charged Nariai solutions at the
point C, corresponding to the critical charge $Q_{\rm C}$. Maximal
black holes with higher charge are stable.}
\label{fig-ss-rnds}
\end{figure}
The Reissner-Nordstr\"om-de~Sitter solution space is shown in
Fig.~\ref{fig-ss-rnds}.  It is a roughly triangular region in the
charge-mass plane. The square of the charge is obviously bounded from
below by zero; this boundary corresponds to the neutral,
Schwarzschild-de~Sitter solutions. From above, it is bounded by the
extremal (or ``cold'') Reissner-Nordstr\"om-de~Sitter solutions. These
solutions also form a lower bound on the mass at any given charge. As
the mass is increased at fixed charge, the black hole radius, $r_2$,
grows, until the black hole is as large as the cosmological horizon
that ``surrounds'' it. This bound corresponds to the Charged Nariai
solutions. The usual Reissner-Nordstr\"om-de~Sitter metric,
Eq.~(\ref{eq-rnds}), breaks down in this limit.  As this paper is
concerned mainly with the peculiar stability properties of the Charged
Nariai solutions, their metric will be given explicitly below.

\subsection{Charged Nariai solution}
  \label{sec-cn}

The Charged Nariai solutions are the black holes of maximal mass for a
given charge. Geometrically, they are obtained in the limit as the
outer black hole horizon radius approaches the radius of the
cosmological horizon. Therefore $r$ will no longer be a suitable
coordinate to describe the region between the two horizons. Instead,
one may set $r_2 = r_0 - \epsilon$, $r_3 = r_0 + \epsilon$. With the
transformation
\begin{equation}
r = r_0 + \epsilon \cos\chi,\ t = \frac{1}{V(r_0)} \epsilon \psi,
\label{eq-limit-rnds}.
\end{equation}
the Reissner-Nordstr\"om-de~Sitter metric, Eq.~(\ref{eq-rnds}),
becomes the Charged Nariai metric,
\begin{equation}
ds^2 = \frac{1}{A} \left( - \sin^2\!\chi d\psi^2 + d\chi^2 \right)
     + \frac{1}{B} d\Omega_2^2,
  \label{eq-CN-metric}
\end{equation}
in the limit $\epsilon \rightarrow 0$. (This limiting procedure is
discussed in more detail, for the neutral case, in the appendix of
Ref.~\cite{BouHaw96}.) Here $A$ and $B$ are given by
\begin{equation}
A = \lim_{\epsilon \rightarrow 0} \frac{V(r_0)}{\epsilon^2},\
B = \lim_{\epsilon \rightarrow 0} \frac{1}{r_0^2}.
\end{equation}
The Maxwell field is still given by Eq.~(\ref{eq-max-magnetic}), and
therefore $F^2 = 2 B^2 Q^2$ in the magnetic case. The parameters $A$
and $B$ are given by
\begin{equation}
A = \frac{1}{2Q^2} \left( 1 - \sqrt{1 - 4 \Lambda Q^2} \right)
,\ \ B = 2 \Lambda - A
. \label{eq-AB-classical}
\end{equation}
Therefore $A<B$ except in the neutral case, when $A=B$. The geometry
can be visualized as a direct product of 1+1-dimensional de~Sitter
space with a two-sphere.

\begin{figure}[htb!]
  \hspace{.1\textwidth} \vbox{\epsfxsize=.8\textwidth
  \epsfbox{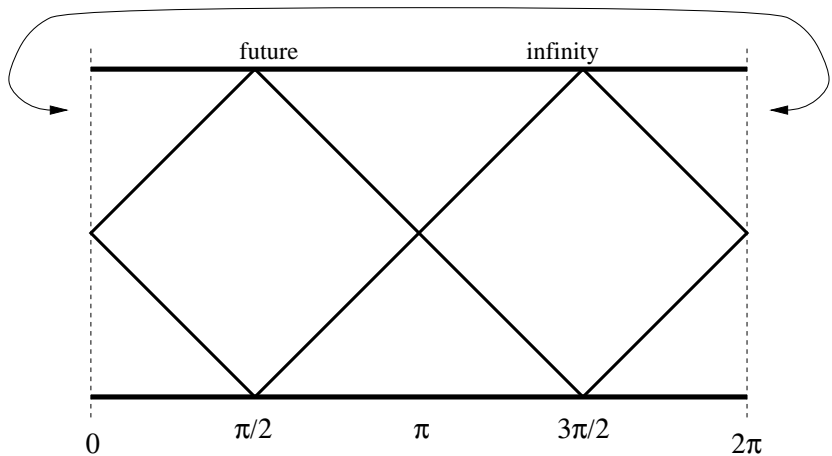}}
\caption%
{\small\sl Penrose diagram for a Charged Nariai solution. It is
  identical to the diagram for 1+1-dimensional de~Sitter space, except
  that every point represents a two-sphere of equal radius,
  $B^{-1/2}$. There is no black hole. Instead, any observer will see
  two cosmological horizons: one in each direction on the $S^1$. They
  are shown here for observers at $\pi/2$ and $3\pi/2$. The solution
  is classically unstable to perturbations of the two-sphere size.
  This causes two-spheres to collapse into black hole interiors or
  grow into de~Sitter regions. If $n$ black holes are formed, the
  upper part of the diagram will be a sequence of $n$
  Reissner-Nordstr\"om-de~Sitter diagrams (Fig.~\ref{fig-cp-rnds-tr}).
  This is shown in Fig.~\ref{fig-cp-prolsub}.}
\label{fig-cp-nariai}
\end{figure}

Note that $0 \leq Q^2 < (4 \Lambda)^{-1} \equiv Q^2_{\rm max}$. At the
upper bound the inner and outer black hole horizons would coincide,
but the metric will be different from
Eq.~(\ref{eq-AB-classical})~\cite{Rom92,ManRos95}. Thus all Charged
Nariai solutions are subextremal; they do, however, closely approach
extremality as $Q^2 \rightarrow Q^2_{\rm max}$. Their causal structure
is shown in Fig.~\ref{fig-cp-nariai}.

For the decay of Charged Nariai black holes to have an effect on the
global structure of de~Sitter space, they must first be produced. This
occurs through gravitational tunneling and can be described using
instantons, or Euclidean solutions of the Einstein-Maxwell
equations~\cite{GinPer83,ManRos95,BouHaw95,BouHaw96,BouCha97}.  Some
issues specific to compact, disconnected instantons are discussed in
Refs.~\cite{BouCha97} and \cite{BraBri98}. For general values of $Q$
and $\mu$, the Reissner-Nordstr\"om-de~Sitter metric,
Eq.~(\ref{eq-rnds}), has no regular Euclidean section. Unless special
boundary conditions are selected~\cite{BouHaw98}, such black holes
cannot nucleate semiclassically on a de~Sitter background. A smooth
instanton does exist, however, for the Charged Nariai solutions (as
well as for the cold, and the `lukewarm' solutions, which will be
discussed later)~\cite{MelMos89,MelMos90,Rom92,ManRos95,BouHaw96}.

The Euclidean Charged Nariai solution can be obtained
straightforwardly by Wick-rotation: With $\xi = i \psi$,
Eq.~(\ref{eq-CN-metric}) describes a Euclidean metric corresponding to
the direct product of two round two-spheres of radii $A^{-1/2}$ and
$B^{-1/2}$. This solution has a Euclidean action of $-2\pi/B$. In
order to obtain the rate at which Charged Nariai black holes are
produced in de~Sitter space, this action must be normalized by
subtracting off the de~Sitter
action~\cite{Col77,GinPer83,BouHaw95,BouCha97}.

The Euclidean de~Sitter solution is given by Eq.~(\ref{eq-rnds}), with
$\mu=Q=0$ and $\tau = it$. This yields a Euclidean four-sphere of
radius $(\Lambda/3)^{-1/2}$. Its action is $-3\pi/\Lambda$; therefore
the Charged Nariai nucleation rate, neglecting a prefactor, is given
by
\begin{equation}
\exp\left( - \frac{\pi}{B}\, \frac{1+2Q^2 B}{1-Q^2 B}  \right).
\end{equation}
This is less than one, and in fact ranges from $\exp(-\pi/\Lambda)$ to
$\exp(-2\pi/\Lambda)$ as the charge increases from $0$ to $Q^2_{\rm
max}$. Therefore, black hole production is suppressed, and more
suppressed the higher the charge. The suppression is weak if the
cosmological constant is close to the Planck value. If it is smaller,
the suppression becomes huge. In an eternal de~Sitter space, however,
the process is bound to occur nevertheless.

\sect{Preliminary stability analysis}
\label{sec-prelim}

The aim of this section is to gain an understanding of the type of
instabilities that may be expected in a Charged Nariai
geometry. Because the spacelike sections are arbitrarily similar to
the submaximal Reissner-Nordstr\"om-de~Sitter solutions, the metric is
classically unstable to the formation of black hole interiors and
large de~Sitter regions. The resulting nearly maximal
Reissner-Nordstr\"om-de~Sitter black holes are classically
stable. However, thermodynamic considerations suggest that some (but
not all) of them will decay by emitting Hawking radiation.

In Sec.~\ref{sec-solutions} these expectations will be verified, and
additional instabilities found, by considering a variety of metric
perturbations in a model that explicitly includes quantum radiation
and back-reaction.

\subsection{Classical instability}
\label{sec-prelim-cl}

The limit leading to Eq.~(\ref{eq-CN-metric}) has made the $S^1 \times
S^2$ topology of the spacelike sections of
Reissner-Nordstr\"om-de~Sitter spaces manifest; $\psi$ is the
coordinate along the one-sphere.  The Charged Nariai solutions are
precisely those for which a slicing can be found in which the
two-sphere radius, $r$, is independent of the $S^1$-coordinate, $x$.
For other Reissner-Nordstr\"om-de~Sitter black holes, the $S^2$ size
varies around the $S^1$; one can find a slicing in which the maximal
two-sphere corresponds to the cosmological horizon, and the minimal
two-sphere to the outer black hole horizon. In this sense, it will be
helpful to think of the spacelike section of a Charged Nariai solution
as a `perfect doughnut,' while it would be a `wobbly doughnut' for a
generic Reissner-Nordstr\"om-de~Sitter solution.

In terms of its space-time geometry, the Charged Nariai solution is
the direct product of 1+1-dimensional de~Sitter space with a round
two-sphere. There is no black hole interior; instead the black hole
horizon has become a second cosmological horizon. Looking in either
direction of the $S^1$, an observer will see a cosmological horizon.
This is shown in the Penrose diagram, Fig.~\ref{fig-cp-nariai}.
However, the Charged Nariai solutions are classically unstable. Any
region in which the two-spheres are slightly smaller will collapse to
form the interior of a black hole. Larger two-spheres will grow
exponentially to form asymptotically de~Sitter regions.

A simple way to understand this instability is to remember that the
Charged Nariai solutions form a set of measure zero in the
Reissner-Nordstr\"om-de~Sitter solution space; yet they have a
completely different causal structure.  For all $\epsilon$, except
$\epsilon=0$, the coordinates given in Eq.~(\ref{eq-limit-rnds})
describe a generic Reissner-Nordstr\"om-de~Sitter solution with the
causal structure of Fig.~\ref{fig-cp-rnds-tr}. Even if at a given
moment a spacelike section has $\epsilon=0$ exactly, a small
$\epsilon$-perturbation would destroy the degeneracy and lead to
Reissner-Nordstr\"om-de~Sitter space.

This perturbation is really the only one allowed by the classical
constraint equations. When quantum matter will be included in
Sec.~\ref{sec-model}, however, more general perturbations with
multiple minima and maxima will become possible (a `doughnut with many
wobbles'). In Sec.~\ref{sec-solutions} I will show that this leads to
necklace configurations: Along the $S^1$, the Charged Nariai spacelike
section will develop a number of asymptotically de~Sitter regions
separated by black hole interiors.

\subsection{Thermodynamics}
\label{sec-prelim-th}

Gibbons and Hawking have shown that the cosmological horizon emits
thermal radiation in de~Sitter space~\cite{GibHaw77a}. Its temperature
is given by the horizon's surface gravity, over $2\pi$, as it would be
for black holes.  Indeed, if a black hole is present, it will exchange
radiation with the cosmological horizon. As a result, it will grow or
shrink, depending on the net influx.

In the particular case of Charged Nariai solutions, the two horizons
are of equal temperature, and will be in thermodynamic equilibrium.
One would not expect this equilibrium to be stable. As discussed
above, a small perturbation will turn the Charged Nariai solution into
a submaximal Reissner-Nordstr\"om-de~Sitter geometry. For the
uncharged case, the following argument was given in
Ref.~\cite{BouHaw97b}: In the perturbed geometry, the black hole will
be slightly smaller than the cosmological horizon. Correspondingly, it
will be hotter, and it will lose more radiation than it absorbs.  Thus
it will continue to shrink at an ever-increasing rate.

One would expect the Charged Nariai solutions to contain a similar
quantum instability. There are some important differences, however.
For example, the black hole mass cannot become smaller than the
extremal value.  Below, the thermodynamic argument is re-examined for
the charged case.  It will turn out that for sufficient charge, black
holes actually anti-evaporate, asymptotically approaching the Charged
Nariai size.

In flat space, a charged black hole does indeed evaporate until it
becomes extremal. In the early stages of evaporation, the charge will
be negligible, and the temperature will increase as the black hole
loses mass. In the extremal limit, however, the temperature approaches
zero and evaporation ceases. Therefore there is a turnaround point: As
the black hole evaporates, the temperature first increases, then
decreases to zero.

The difference in de Sitter space is that the black hole cannot become
arbitrarily cold. As it approaches the extremal limit, its temperature
eventually drops to that of the cosmological horizon. At this level
the black hole will be stabilized by the Gibbons-Hawking radiation
bath it is immersed in. Indeed there are two families within the
Reissner-Nordstr\"om-de~Sitter solution space for which the black hole
is in thermodynamic equilibrium with the cosmological horizon. The
first is the unstable Charged Nariai family, for which the black holes
are large. The second is the family of `lukewarm'
solutions~\cite{MelMos89}, which are characterized by the condition $Q^2
= \mu^2$ in the metric, Eq.~(\ref{eq-rnds}). As shown in
Fig.~\ref{fig-ss-rnds}, the lukewarm solutions meet the Charged Nariai
solutions at the point C, corresponding to a critical charge $Q^2_{\rm
C} = 3/(16\Lambda) = 3 Q^2_{\rm max} / 4$.

First consider black holes with subcritical charge ($Q^2 < Q^2_{\rm
  C}$). If $Q^2<\mu^2$, the black hole is hotter than the cosmological
horizon, and will evaporate until it becomes a lukewarm solution. This
includes the case of a perturbed Charged Nariai solution. Subcritical
black holes with $Q^2>\mu^2$ are colder than the lukewarm solution and
will absorb radiation from the cosmological horizon until they reach
the lukewarm level. Thus the lukewarm black holes form the stable
endpoints of the evolution of subcritically charged black holes.

Now consider the black holes of supercritical charge ($Q^2 > Q^2_{\rm
C}$). There are no equilibrium solutions between the Charged Nariai
solutions, in which the black hole is as hot as the cosmological
horizon, and the cold solutions, in which the black hole temperature
(but not the cosmological temperature) vanishes. Therefore the black
hole is colder than the cosmological horizon for all but the Charged
Nariai solutions. It will absorb cosmological radiation and grow,
asymptotically approaching the Charged Nariai limit.

In summary, all Charged Nariai solutions are expected to be
classically unstable to the formation of a black hole interior. If
$Q^2 < Q^2_{\rm C}$, the black hole is expected to be
thermodynamically unstable; for larger charges, it should be stable.

\sect{Including radiation} \label{sec-model}

\subsection{Model construction}

The production of a Charged Nariai geometry, described in the previous
section, requires only semiclassical techniques. The process can be
thought of as a non-perturbative fluctuation of the gravitational
field, mediated by instantons. To continue with the description of the
fragmentation process, however, it is necessary to work in a model
that includes the quantum radiation exchanged between the black hole
and cosmological horizons.

Black holes in asymptotically flat space have been shown to
radiate~\cite{Haw74}; they will lose mass, shrink, and unless they are
stabilized by charge, they will eventually disappear. For black holes
in de~Sitter space, the consequences of this radiation are less
obvious, since the cosmological horizon also
radiates~\cite{GibHaw77a}.  For the Nariai solution the two horizons
are initially of equal size and temperature, so the black hole suffers
no net loss of energy. In Ref.~\cite{BouHaw97b} we investigated the
stability of this equilibrium, considering only perturbations
corresponding to the formation of a single black hole. For this
purpose, a spherically symmetric model was introduced that included
the one-loop effective action of $N$ scalar fields. The same model was
employed in Ref.~\cite{Bou98} to demonstrate the proliferation effect
arising from higher mode perturbations of the Nariai solution. Here it
will be used to investigate whether the Charged Nariai solutions
contain a similar instability.  (The stability of the neutral Nariai
solution was investigated using more elaborate models in
Refs.~\cite{NojOdi98b,EliNoj99}.)

Restricting to spherically symmetric fields and quantum fluctuations,
the metric may be written as
\begin{equation}
ds^2 = e^{2\rho} \left( -dt^2 + dx^2 \right) + e^{-2\phi} d\Omega^2,
\label{eq-ssans}
\end{equation}
where $x$ is the coordinate on the $S^1$, with period $2\pi$.  Using
this ansatz, and the on-shell condition for magnetic fields,
\begin{equation}
F_{\mu\nu} F^{\mu\nu} = 2 Q^2 e^{4\phi},
\end{equation}
the angular coordinates and the Maxwell field can be integrated out in
Eq.~(\ref{eq-action-4D}), which reduces the action to
\begin{equation}
S = \frac{1}{16\pi} \int d^2\!x\, (-g)^{1/2} e^{-2\phi} \left[
 R + 2 (\nabla \phi)^2 + 2 e^{2\phi} - 2
 \Lambda - 2 Q^2 e^{4\phi} - \frac{1}{2}
 \sum_{i=1}^{N} (\nabla f_i)^2 \right],
\end{equation}

The scalars couple to the dilaton in the two-dimensional action. Thus,
to include the back-reaction from quantum radiation, one should find
the classical solutions to the action $S+W^*$, where $W^*$ is the
scale-dependent part of the one-loop effective action for dilaton
coupled scalars~\cite{EliNaf94,MukWip94,ChiSii97,BouHaw97a,NojOdi97,%
NojOdi98e,Ich97,KumLie97,Dow98}:
\begin{equation}
W^* = - \frac{1}{48\pi} \int d^2\!x (-g)^{1/2} \left[ \frac{1}{2}
  R \frac{1}{\Box} R - 6 (\nabla \phi)^2 \frac{1}{\Box} R
  - w \phi R \right].
\end{equation}
In the large $N$ limit, the contribution from the quantum fluctuations
of the scalars dominates over that from the metric fluctuations.  In
order for quantum corrections to be small, one should take $ N \Lambda
\ll 1 $.  For small perturbations of the charged Nariai solutions, the
$(\nabla \phi)^2$ term may be neglected~\cite{BouHaw97b}. The
coefficient $w$ will not be specified here. It will later drop out of
the calculation; thus, the $\phi R$ term will not affect any results at
the present level of approximation.

One can obtain a local form of this action by introducing an
independent scalar field $Z$ which mimics the trace anomaly. With the
classical solution $f_i=0$ the $N$ scalars can be integrated out. Thus
the action of the one-loop model will be given by:
\begin{eqnarray}
S\!\!\! &
 =\!\!\! & \frac{1}{16\pi} \int d^2\!x\, (-g)^{1/2} \left[ \left(
e^{-2\phi} + \frac{ N }{3} (Z + w \phi) \right) R \right.
\nonumber \\
& & \mbox{\hspace{2em}} \left.
 - \frac{ N }{6} \left( \nabla Z \right)^2
+ 2 + 2 e^{-2\phi} \left( \nabla \phi \right)^2
- 2 e^{-2\phi} \Lambda -2 Q^2 e^{2\phi} \right]\!.
\end{eqnarray}

\subsection{Equations of motion}

Differentiation with respect to $t$ ($x$) will be denoted by an
overdot (a prime). For any functions $f$ and $g$, define:
\begin{equation}
\partial f\,\partial g  \equiv - \dot{f} \dot{g} + f' g',\ \ \ \
\partial^2 g \equiv - \ddot{g} + g'',
\end{equation}
\begin{equation}
\delta f\,\delta g \equiv \dot{f} \dot{g} + f' g',\ \ \ \
\delta^2 g \equiv \ddot{g} + g''.
\end{equation}
Variation with respect to $\rho$, $\phi$ and $Z$ yields the
following equations of motion:
\begin{equation}
- \left( 1 - \frac{wN}{6} e^{2\phi} \right)
 \partial^2 \phi + 2
(\partial \phi)^2 + \frac{N}{6} e^{2\phi} \partial^2 Z +
e^{2\rho+2\phi} \left( \Lambda e^{-2\phi}
+ Q^2 e^{2\phi} - 1 \right) = 0;
\label{eq-m-rho}
\end{equation}
\begin{equation}
\left( 1 - \frac{wN}{6} e^{2\phi}
 \right) \partial^2 \rho -
\partial^2 \phi + (\partial \phi)^2 + \Lambda e^{2\rho}
- Q^2 e^{2\rho+4\phi} = 0;
\label{eq-m-phi}
\end{equation}
\begin{equation}
\partial^2 Z - 2 \partial^2 \rho = 0.
\label{eq-m-Z}
\end{equation}
The constraint equations are:
\begin{equation}
\left( 1 - \frac{wN}{6} e^{2\phi} \right)
 \left( \delta^2 \phi - 2 \delta\phi\,\delta\rho \right) -
(\delta\phi)^2
= \frac{N}{12} e^{2\phi} \left[ (\delta Z)^2 + 2 \delta^2 Z
  - 4 \delta Z \delta\rho \right];
\label{eq-c1}
\end{equation}
\begin{equation}
\left( 1 - \frac{wN}{6} e^{2\phi} \right)
 \left( \dot{\phi}' -
\dot{\rho} \phi' - \rho' \dot{\phi} \right) - \dot{\phi} \phi'
= \frac{N}{12} e^{2\phi} \left[ \dot{Z} Z' + 2
\dot{Z}' - 2 \left( \dot{\rho} Z' + \rho'
  \dot{Z} \right) \right].
\label{eq-c2}
\end{equation}
From Eq.~(\ref{eq-m-Z}), it follows that $Z = 2\rho + \eta$, where
$\eta$ satisfies $\partial^2 \eta = 0$. The remaining freedom in
$\eta$ can be used to satisfy the constraint equations for any choice
of $\rho$, $ \dot{\rho} $, $\phi$ and $\dot{\phi}$ on an initial
spacelike section~\cite{BouHaw97b}.

\sect{Linear perturbations and back-reaction} \label{sec-solutions}

\subsection{Quantum Charged Nariai solution}

Using the model established in the previous section, the stability of
the Charged Nariai solutions can now be investigated. The analysis
will follow the procedure set forth (for neutral black holes) in
Refs.~\cite{BouHaw97b,Bou98}, where more details can be found. Here
the emphasis will be on the effects of charge.

The Charged Nariai solution, Eq.~(\ref{eq-CN-metric}), may be
rewritten in the form of Eq.~(\ref{eq-ssans}). This yields
\begin{equation}
e^{2\rho} = \frac{1}{A} \frac{1}{\cos^2\! t}, \ \ \
e^{2\phi} = B,
\label{eq-CN-metric2}
\end{equation}
with $A$ and $B$ given by Eq.~(\ref{eq-AB-classical}).

The next step is to find the quantum corrections to the unperturbed
solution. They take the form of small corrections to $A$ and $B$, and
can be obtained by substituting Eq.~(\ref{eq-CN-metric2}) into the
one-loop equations of motion, Eqs.~(\ref{eq-m-rho}) and
(\ref{eq-m-phi}). This yields the following cubic equation for $B$:
\begin{equation} 
B \left[ \left( 1 - \frac{wN}{6} B \right)
- Q^2 B \left( 1 - \frac{(w-2)N}{6} B \right) \right]
= \left[ 1 - \frac{(w+2)N}{6} B \right] \Lambda,
\label{eq-B}
\end{equation}
while $A$ is given by
\begin{equation}
A = \frac{\Lambda - Q^2 B^2}{1 - wNB/6}.
\end{equation}
Expanding in the small parameter $N \Lambda$, one obtains to first
order:
\begin{eqnarray} 
A & = & A_0 \left[ 1 + \frac{N B_0}{3} \left( \frac{2 Q^2 B_0}{1 -
        2 Q^2 B_0} + \frac{w}{2} \right) \right],\\
B & = & B_0 \left( 1 - \frac{N B_0}{3} \right),
\end{eqnarray}
where $A_0$ and $B_0$ refer to the classical solution,
Eq.~(\ref{eq-AB-classical}).  Note that in the near-extremal limit,
one has $2 Q^2 B_0 \rightarrow 1$, and the expansion in $N \Lambda$
breaks down.

\subsection{Metric perturbation}

Quantum fluctuations will destroy the degeneracy of the Charged Nariai
solution. This means that the two-sphere radius, $e^{-\phi}$, will no
longer be constant. Instead, it will be a function of the one-sphere
coordinate, $x$, and time, $t$. It can be decomposed into Fourier
modes with time-dependent coefficients. It simplifies the analysis
considerably to assume that one mode dominates over all others; in
this case the black holes will be distributed evenly around the
one-sphere. The probability for the $n$-th mode to dominate will be of
the order of $\exp(-n^2)$. The corresponding perturbation may be
written as
\begin{equation}
e^{2\phi} =  B \left[ 1 + 2 \epsilon \sigma_n(t) \cos nx \right],
\label{eq-pert-phi}
\end{equation}

This simple ansatz is sufficient to illustrate various evolutionary
possibilities for black hole-de~Sitter solutions which ultimately
affect the global structure of de~Sitter space, including the
formation of one or several charged or uncharged black holes on a
necklace, their evaporation or anti-evaporation, and their
fragmentation.  Nevertheless, it has limitations.  Most importantly,
the perturbation can be considered linear only as long as the
two-sphere size does not differ significantly from the Charged Nariai
value anywhere on the $S^1$. Yet the perturbations soon become
non-linear in the black hole interiors, where the two-spheres collapse
to zero size, and in the de~Sitter regions, where they expand
indefinitely. Even considering only the regions between the black hole
and cosmological horizon, the linear approximation eventually breaks
down, if the black hole evaporates. (This is only a problem if its
charge is small; otherwise a large lukewarm endpoint will preclude
further evaporation, or, for supercritical charge, evaporation never
begins.) It is fair to assume, of course, that a very small black hole
will hardly notice the cosmological horizon and can be treated like a
Schwarzschild black hole. Numerical work is currently underway to
verify the smooth transition between the perturbative and
non-perturbative regimes~\cite{BouNie99}. We will also investigate
more general perturbations and examine the conditions for domination
of certain modes.

I will not consider the constant mode perturbation ($n=0$) in detail,
but most of the analysis for general $n$ will apply to it. If the
two-spheres are all larger than the Charged Nariai value, they will
expand exponentially everywhere. The resulting spacetime will be
locally de~Sitter, with no black holes and a connected future
infinity. Globally its topology will be $S^1 \times S^2$. If the
two-spheres are smaller, they will collapse to a singularity
everywhere, unless they are stabilized by quantum radiation at a small
radius. It is not clear whether this should be interpreted as the
collapse of the entire de~Sitter space; related questions are
discussed in Sec.~\ref{sec-discussion}.

In order to obtain an equation of motion for the metric perturbation,
$\sigma_n$, one may eliminate $\partial^2 \rho$ and $\partial^2 Z$
from Eq.~(\ref{eq-m-rho}). This yields
\[ 
- \left[ 1 - (w+1) \frac{N}{3} e^{2\phi} + \left(\frac{wN}{6}
  e^{2\phi} \right)^2 \right] \partial^2 \phi
+ \left[ 2 - (w+1) \frac{N}{3} e^{2\phi} \right] (\partial \phi)^2 =
\]
\begin{equation}
= e^{2\rho} \left\{
\left( 1 - \frac{wN}{6} e^{2\phi} \right) e^{2\phi}
- \left[ 1 - (w-2) \frac{N}{6} e^{2\phi} \right] Q^2 e^{4\phi}
- \left[ 1 - (w+2) \frac{N}{6} e^{2\phi} \right] \Lambda \right\}.
\label{eq-phi}
\end{equation}
Note that the left- and right-hand sides of this equation vanish
separately in the unperturbed case ($\epsilon=0$). Therefore,
perturbations of $\rho$ will drop out at first order in $\epsilon$ and
will not be considered here.  Insertion of the perturbation ansatz,
Eq.~(\ref{eq-pert-phi}), yields
\begin{equation}
\frac{\ddot{\sigma}_n}{\sigma_n} = \frac{a}{\cos^2\! t} - n^2,
\label{eq-m-si}
\end{equation}
where
\begin{equation} 
a = \frac{2 \left[ 6 \!-\! 2 wNB \!-\!
12 Q^2 B \!+\! 3 (w\!-\!2)N Q^2 B^2 \!+\! (w\!+\!2) N
\Lambda \right] (6 \!-\! wNB) B}{\left[ 36 \!-\!
12 (w\!+\!1) NB \!+\! w^2 N^2 B^2
\right] (\Lambda \!-\! Q^2 B^2)}.
\label{eq-a-full}
\end{equation}
\begin{figure}[htb!]
  \hspace{.2\textwidth} \vbox{\epsfxsize=.6\textwidth
  \epsfbox{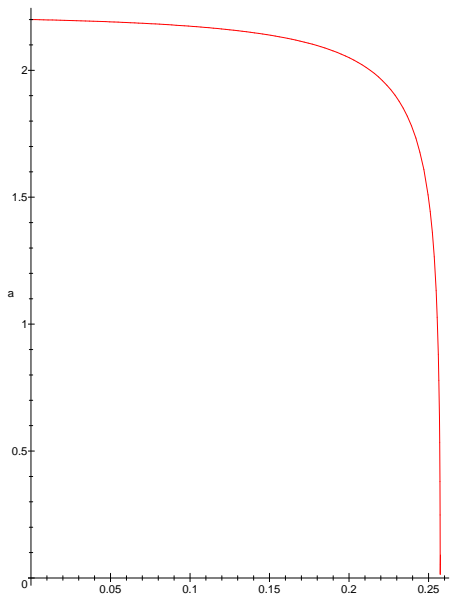}}
\caption%
{\small\sl A typical plot for the parameter $a$ as a function of $Q^2
\Lambda$.  $a>2$ for small charges. $a$ passes through 2 at the
critical charge, near $Q^2 \Lambda=3/16$. Black holes with larger
charge will anti-evaporate. $a$ becomes zero exactly in the limit
where the Charged Nariai solutions become extremal, near $Q^2 \Lambda
= 1/4$.}
\label{fig-avsQ}
\end{figure}

The value of $a$ is crucial in determining whether a black hole grows
or shrinks; in the neutral case it can be expressed in terms of
$\Lambda$, $N$, and $w$ only~\cite{BouHaw97b}. In order to evaluate
$a$ for $Q^2>0$, however, one would need to solve the cubic equation
(\ref{eq-B}) for $B$. It seems preferable to evaluate $a$ only to
first order in $N \Lambda$. This gives a surprisingly simple result:
\begin{equation}
a = 2 + \left( \frac{2 N B_0}{3} \right)
    \left( \frac{1 - 4 Q^2 B_0}{1 - 2 Q^2 B_0} \right).
\label{eq-a}
\end{equation}

As advertised, this expression is independent of $w$.  The first
factor will be small as long as $N \Lambda \ll 1$, which must be
assumed anyway for the one-loop effective action to make sense. The
second factor will be of the order of $1$, except in the near-extremal
limit, when $2 Q^2 B_0 \approx 1$. There the expansion in $N \Lambda$
cannot be trusted, and should be supplemented by a numerical
calculation of $a$. Fig.~\ref{fig-avsQ} shows a plot of $a$ as a
function of the black hole charge.
Note that $4 Q^2 B_0 < 1$ for $Q < Q^2_{\rm C} = 3/(16\Lambda)$, and
$4 Q^2 B_0 > 1$ for $Q > Q^2_{\rm C}$. Therefore, $a$ will be greater
(less) than 2 for subcritical (supercritical) black hole charge.

\subsection{Horizon perturbation}

The metric perturbations considered above lead to the formation of
black hole and cosmological horizons.  The condition for an apparent
horizon is $(\nabla \phi)^2 = 0$. Eq.~(\ref{eq-pert-phi}) yields
\begin{equation}
\frac{\partial\phi}{\partial t} = \epsilon\, \dot{\sigma}_n
 \cos nx,\ \ \
\frac{\partial\phi}{\partial x} = - \epsilon\, \sigma_n\, n \sin nx.
\end{equation}
Therefore, there will be $2n$ black hole horizons, and $2n$
cosmological horizons, located at
\begin{eqnarray}
 x_{{\rm b}}^{(k)}(t) & = &
 \frac{1}{n} \left( 2\pi k + \arctan \left|
 \frac{\dot{\sigma}_n}{n\sigma_n} \right| \right), \\
 x_{{\rm b}}^{(n+k)}(t) & = &
 \frac{1}{n} \left( - 2\pi k + \arctan \left|
 \frac{\dot{\sigma}_n}{n\sigma_n} \right| \right), \\
 x_{{\rm c}}^{(l)}(t) & = & x_{{\rm b}}^{(l)}(t) + \frac{\pi}{n},
\label{eq-chib}
\end{eqnarray}
where $k = 0 \ldots n-1$ and $l = 0 \ldots 2n-1$.

By inserting these values of $x$ back into the metric perturbation,
Eq.~(\ref{eq-pert-phi}), one finds the size of the black hole
horizons:
\begin{equation}
r_{{\rm b}}(t)^{-2} = e^{2\phi[t, x^{(l)}_{{\rm b}}(t)]} =
B \left[ 1 + 2 \epsilon \delta(t) \right],
\label{eq-rb}
\end{equation}
where
\begin{equation}
\delta \equiv \sigma_n \cos n x^{(l)}_{\rm b} =
\sigma_n \left( 1+
 \frac{\dot{\sigma_n}^2}{n^2\sigma_n^2} \right)^{-1/2}.
\label{eq-delta}
\end{equation}
will be called the {\em horizon perturbation}.

If $\delta$ grows, the black hole is shrinking; this corresponds to
evaporation. To see whether this happens, one needs to find solutions
for $\sigma_n$ for given initial conditions, and plug them into
Eq.~(\ref{eq-delta}).

\subsection{Early time evolution}

Exact solutions to Eq.~(\ref{eq-m-si}) have not been found for general
values of $a$. Various approximations can be used, however, to
investigate the initial behavior of the black holes, and to determine
whether they ultimately evaporate, or grow.

The possible initial conditions for $\sigma_n$ can be parametrized by
writing:
\begin{equation}
\sigma_n(0) = \alpha \sin \vartheta,\ \ \ \
\dot{\sigma}_n(0) = \alpha \cos \vartheta.
\end{equation}
To obtain the early-time behavior of the horizons, one may solve the
equation of motion, Eq.~(\ref{eq-m-si}), for a power series in $t$,
and use the result in the equation for the horizon perturbation,
Eq.~(\ref{eq-delta}). Writing down the general expressions for its
coefficients would not be very illuminating. It suffices to say that
the coefficients $\delta_i$ can be positive or negative, depending on
$a$, $\vartheta$, and $n$, and not all of them will have the same sign
generically.

This makes the early-time behavior quite complicated. The main point,
however, is that black holes with $Q<Q_{\rm C}$, which are
thermodynamically unstable, will not necessarily evaporate
initially. And supercritically charged black holes, which ought to be
stable according to Sec.~\ref{sec-prelim-th}, may initially
become smaller. This can be demonstrated by choosing $\vartheta =
\pi/2$ and $n=1$. (Note that this choice corresponds to vanishing
time-dependence. It thus gives the initial conditions most similar to
a slightly submaximal Reissner-Nordstr\"om-de~Sitter black hole.)

For $\vartheta = \pi/2$, the linear coefficient, $\delta_1$, vanishes.
The early time evolution is determined by the quadratic coefficient,
and is given by
\begin{equation}
\delta(t) = \alpha \left[ 1 - \frac{1}{2} (a-1) (a-2) t^2 \right].
\end{equation}
By Eq.~(\ref{eq-a}), $a>2$ for black holes of subcritical charge. Thus
the horizon perturbation begins to decrease. Therefore, contrary to
thermodynamic expectations, such black holes will initially grow, or
`anti-evaporate'. On the other hand, if the charge is supercritical
(but sufficiently non-extremal), $a$ will lie between 1 and 2, and the
horizon perturbations begin to increase. These black holes evaporate
initially.

This behavior may be counterintuitive, but it is not obviously absurd.
In de~Sitter space, the quantum radiation is distributed on a compact
spatial manifold. Through the constraint equations, (\ref{eq-c1}) and
(\ref{eq-c2}), the initial metric perturbation completely determines
the distribution of the quantum radiation around the $S^1$. This
distribution may correspond to energy heading towards the black hole
even when it is thermodynamically hotter, and vice versa. This does
not violate the Second Law. It means only that the radiation field is
initially out of equilibrium with the horizons.

But what happens at later times? Does the radiation field relax and
the thermodynamic evolution prevail?  To answer this question fully a
numerical study is called for, work on which is
underway~\cite{BouNie99}. Analytically, two things can be done. First,
I will give asymptotic solutions to the linear perturbation equation
at late times. Second, I will investigate the particular initial
conditions selected whenever the Nariai geometry emerges from a
semiclassical transition.

\subsection{Late time evolution}

It will be useful to rescale the time variable in
Eq.~(\ref{eq-CN-metric}) as $\cosh v = 1/\cos t$. Then the Nariai metric
takes the form
\begin{equation}
ds^2 =  \frac{1}{A} \left( - dv^2 +
  \cosh^2\! v\, d x^2 \right) +  \frac{1}{B} d\Omega^2.
\label{eq-nariai-v}
\end{equation}
In the new time coordinate, the metric perturbation equation
(\ref{eq-m-si}) becomes
\begin{equation}
\frac{d^2 \sigma_n}{dv^2} + \tanh v \frac{d \sigma_n}{dv}
 + \left( \frac{n^2}{\cosh^2 v} - a \right) \sigma_n = 0.
\end{equation}

At late times ($v \gg 1$), one can take $\tanh v \approx 1$ and
neglect the $n^2$ term. Asymptotic solutions are therefore given by
\begin{equation}
\sigma_n(v) = \alpha_+ \exp(c_+ v) + \alpha_- \exp(c_- v),
\label{eq-sigma-late}
\end{equation}
where $\alpha_\pm$ are arbitrary constants and $c_\pm$ are the two
solutions to $c(c+1)=a$. Unless $\alpha_+$ is exactly 0, which would
be unphysical fine-tuning, the mode with the larger exponent, $c_+$,
will eventually dominate no matter how small its initial excitation.

Neglecting the weaker mode, the horizon perturbation can be found from
Eq.~(\ref{eq-delta}):
\begin{equation}
\delta = 2 \alpha_+ n c_+^{-1} \exp \left[ (c_+ - 1) v \right].
\label{eq-delta-late}
\end{equation}
The exponent is positive (negative) for $c_+ > 1$ ($c_+ < 1$). Thus
black holes will evaporate if $a > 2$, and they will anti-evaporate if
$a < 2$. But by Eq.~(\ref{eq-a}), $a$ goes through 2 precisely when
the charge becomes critical. This means that Charged Nariai black
holes will evolve according to thermodynamic expectations at late
times, even if they initially fail to do so. The conclusion holds for
generic $\vartheta$ and all $n$. Therefore, the anti-evaporation
effect found for neutral black holes in Ref.~\cite{BouHaw97b} is
transitory, and eventually gives way to evaporation. Similarly, the
evaporation found for supercritically charged black holes will give
way to growth.\footnote{Before this argument was found, the black hole
`turnaround' was first demonstrated numerically by Jens
Niemeyer~\cite{BouNie99}.}

Equation (\ref{eq-a}) and the full expression for $a$,
Eq.~(\ref{eq-a-full}), arise from complicated combinations of one-loop
terms in the equations of motion. Yet they give perfect agreement with
the thermodynamic analysis, and even predict the same critical charge
beyond which evaporation is replaced by black hole growth. This is a
non-trivial check. It shows that this two-dimensional effective model
is capable of accurately reproducing the thermodynamic properties of
the four-dimensional solutions. In addition, of course, it also
incorporates back-reaction effects.

\subsection{Euclidean boundary conditions} \label{sec-regsolutions}

So far, I have considered general linear perturbations of the Charged
Nariai geometry. When this solution nucleates by semiclassical
tunneling, however, it will emerge from a compact Euclidean
solution. Any perturbation has to be regular in the entire Euclidean
region. Otherwise the geometry would not be a solution. It would fail
to dominate the path integral, and would be semiclassically
forbidden.

The regularity condition effectively selects a particular value of the
phase, $\vartheta$, on the $t=0$ surface. The perturbation amplitude
can still be chosen freely, but the relative strength of $\sigma_n$
and $\dot{\sigma}_n$ is fixed. In Ref.~\cite{Bou98}, an approximate
solution to Eq.~(\ref{eq-m-si}) was found which is everywhere regular
and extends far beyond the nucleation surface, so that the late-time
evolution of the black hole can be seen. The solution depends only on
$a$ and $n$, and was found for general $a$, so that only a brief
review will be given here.

Wick-rotation, $u = \frac{\pi}{2} + iv$, yields the Euclidean version
of the Charged Nariai metric, Eq.~(\ref{eq-nariai-v}):
\begin{equation}
ds^2 =  \frac{1}{A} \left( du^2 +
  \sin^2\! u\, d x^2 \right) +  \frac{1}{B} d\Omega^2.
\label{eq-nariai-u}
\end{equation}
This $S^2 \times S^2$ instanton describes the spontaneous nucleation
of a degenerate handle in de~Sitter space.  A suitable nucleation path
runs from the South pole of the first two-sphere, at $u=0$, to
$u=\pi/2$, and then parallel to the imaginary time axis ($u=\pi/2+iv$)
from $v=0$ to $v=\infty$. Geometrically, this corresponds to cutting
the first two-sphere in half, and joining to it a Lorentzian
1+1-dimensional de~Sitter hyperboloid.

On the South pole, the $S^1$ factor of the spacelike sections
degenerates into a point. Therefore, the perturbation may not depend
on the angular variable $x$ there. This means that the condition
$\sigma_n(u=0)=0$ must be imposed when Eq.~(\ref{eq-m-si}) is solved.
For any $n \geq 2$, the family of solutions,
parametrized by a real prefactor $\alpha$, is given by~\cite{Bou98}
\begin{equation}
\sigma_n(u) = 2 \alpha e^{i(c_+ - n) \pi/2} \left( \tan \frac{u}{2}
  \right)^n \left( n + \cos c_+ u \right);
\end{equation}
for $n=1$, it is~\cite{BouHaw97b}
\begin{equation}
\sigma_1(u) = 2 \alpha e^{i(c_+ -1) \pi/2} \sin c_+ u.
\end{equation}
The choice of phase ensures that $\sigma_n$ will be real at late
Lorentzian times, when measurements are made. As before, $c_+$ is the
larger root of $c(c+1)=a$.  The solutions are exact only for $c_+ =1$
(which corresponds to no quantum matter) but are a good approximation
for the whole range of $c_+ $ attained by charged black holes.

This solution is particularly useful because it describes the entire
evolution, from the creation of the Charged Nariai geometry to the
evaporation of the black holes, in a single expression. For small
Lorentzian times, $0 \leq v \leq \mbox{arsinh}\, n$, the metric
perturbation is contained within a single Hubble volume, and
oscillates. When the $S^1$ size has expanded by a factor of $n$, at $
v \approx \mbox{arsinh}\, n$, the perturbations leave the horizon and
freeze out. The $n$ maxima run away to form asymptotically de~Sitter
regions, while the $n$ minima collapse to form black hole interiors.
Thus a necklace configuration develops.  For larger $v$, the solution
asymptotes to
\begin{equation}
\sigma_n(v) = \alpha \exp (c_+ v).
\end{equation}
By Eq.~(\ref{eq-sigma-late}), this agrees with the late time attractor
found above, as it should.

A discussion of the corresponding evolution of the horizon
perturbation has already been given after Eq.~(\ref{eq-delta-late}).
Subcritically charged Nariai solutions become a necklace with $n$
beads, as expected. A finite number of de~Sitter universes results
(see Figs.~\ref{fig-structure} and \ref{fig-cp-prolsub}).
Supercritically charged Nariai solutions also decay into a necklace,
but the black holes do not become small. Instead they anti-evaporate
and approach the Charged Nariai limit. There will be $2n$ regions in
which the two-sphere size is nearly constant across a Hubble length on
the $S^1$, namely each region between a black hole horizon and its
neighboring cosmological horizon. Classically, the geometry is locally
Nariai in such regions. Higher-mode fluctuations continue to push the
two-sphere size over and under the degenerate value locally. As these
perturbations freeze out, more black hole and cosmological regions
will be seeded. Ultimately, this effect leads to the formation of an
infinite number of de~Sitter beads along the same $S^1$ (see
Figs.~\ref{fig-structure} and \ref{fig-cp-prolsup}).

\section{Discussion: Global vs.\ local perspective}
\label{sec-discussion}

In Ref.~\cite{Bou98} I showed that de~Sitter space is unstable to the
proliferation into disconnected daughter universes. This fragmentation
is mediated by spontaneously nucleated neutral black holes. In the
present paper, I have investigated the consequences of charged black
hole creation in de~Sitter space, and found that it leads to the
formation of a `necklace' of de~Sitter universes. Unlike the neutral
case, the necklace does not fragment. Instead, the de~Sitter universes
remain connected by charged black hole throats. The most significant
difference to the neutral case is an iterative effect occuring if the
black holes contain more than a certain critical charge. In this case,
a single nucleation event results in the formation of an unbounded
number of daughter universes.

Sec.~\ref{sec-outline} contains a more detailed summary of the
processes found in this paper.  Here I will discuss some open
questions.

During inflation, the universe was effectively in a de~Sitter
state. In most models, the proliferation effect occurs; this means
that we live in one of a large or infinite number of universes that
originated in the same inflationary region~\cite{Bou98}. But unless
the other universes lead to subtle, non-local quantum effects, we
cannot verify their existence even in principle, because they will
never intersect our causal past. One may ask, therefore, whether
Occam's razor should not be applied to the whole scenario. I believe
this would be a mistake. Occam's razor applies to theories, not to
their solutions. A theory can be simple and successful in describing
every single experiment we can perform, and yet it may predict many
phenomena outside our range of observation. For quantum gravity, this
is actually to be expected. One cannot consider a theory complete, and
yet reject the picture it gives us for the global structure of
spacetime. (Euclidean quantum gravity, of course, is neither complete
nor consistent, but it may give a good semi-classical approximation to
the full theory.)  Otherwise it would also be wrong, for example, to
speak of the `present universe,' or think of it as a homogeneous
constant curvature space; all we can observe, after all, is our past
lightcone.

But what if genuine reasons are found for abandoning the global point
of view in studying de~Sitter space?  Would this render the effects of
fragmentation and proliferation irrelevant?  A local observer cannot
see all of the spacetime; half of it will be hidden behind an event
horizon. If quantum effects are included and de~Sitter space
fragments, the daughter universes will be causally disconnected; a
local observer will never know about more than one of them. She may
notice an endpoint of black hole evaporation; but she has no way of
telling whether the black hole throat really was a connection to a
different de~Sitter region, or whether it merely wrapped around to the
other end of her own, as in the classical
Reissner-Nordstr\"om-de~Sitter solution. Is there any way, then, that
a local observer can find out about the global structure? This
question is one reason why the new process of infinite bead production
in the supercritically charged Nariai geometry is important.  An
observer in a region between a black hole and cosmological horizon
will actually notice the constant production of new daughter
universes.  From her perspective, this corresponds to flipping the
direction in which the black hole lies, i.e., in which the two-sphere
radius decreases.  So at least in the supercritical case, the
fragmentation of future infinity can be inferred even by a local
observer.

The transition from de~Sitter to Charged Nariai is a topology change,
$S^3 \rightarrow S^1 \times S^2$; it is like punching a hole through
the three-sphere. Asymptotic de~Sitter regions develop in the
resulting spacetime. From a local point of view, they are
indistinguishable from the de~Sitter background on which the
transition first happened, so one would expect more topological
transitions to occur in the daughter universes. This argument is
crucial for the production of an infinite number of disconnected
universes in the neutral case (proliferation). But if one tries
explicitly to specify surfaces onto which the de~Sitter instanton can
match, one finds that they necessarily cross the original nucleation
surface, and are therefore globally distinct from a de~Sitter
background\footnote{I am indebted to Ted Jacobson for discussions of
this question.}. The supercritical case of infinite bead production on
the same $S^1$ circumvents this problem, which is another reason why I
consider it significant.

Related questions arise for other instanton-mediated processes in
de~Sitter space, such as the nucleation of cosmic
strings~\cite{GarVil92} or the Hawking-Moss
transition~\cite{HawMos82}.  From a global point of view, there should
only ever be one such transition, occurring at a minimal three-sphere
(the `waist') of de~Sitter space. This is because no minimal
three-sphere background for additional transitions is ever available
thereafter if the global picture is taken at face value.  But from a
local point of view there will be many Hubble volumes in which the
space is asymptotically indistinguishable from the background. It
would seem absurd that the process should not be allowed there.
Similarly, in the Hawking-Moss case, it has been
debated~\cite{HawMos82,HawMos83,Lin98} whether the whole space, or
just one Hubble volume, tunnels into the new state. In pure de~Sitter
space such distinctions may just be meaningless. But for tunneling
events in an inflationary universe the difference is significant.  A
post-inflationary observer will be able to see widely separated
regions of the effective de~Sitter space, and will be able to
determine whether tunneling occured more than once.

As mentioned in the introduction, it has been suggested that past and
future infinity form the holographic surfaces of de~Sitter
space~\cite{Wit98SB}. This conjecture takes a global point of view.
In Ref.~\cite{Bou98} and the present paper I have shown that it would
be inconsistent with quantum theory to assume future infinity to be
connected. If future infinity is indeed a holographic surface, this
result will have consequences for the implementation of the
holographic principle in de~Sitter space.

The conjecture was motivated by the compactness of space, and the lack
of a null infinity, in fully extended de~Sitter space. From a local
point of view, however, one should not consider the whole of de~Sitter
space.  The geometry is effectively bounded by a null surface, the
observer's event horizon (see Fig.~\ref{fig-cp-ds}).  Could this
horizon be a holographic surface? Then the fragmentation of future
infinity would not cause any complications; a local observer can only
reach a single point of future infinity anyway.

\section*{Acknowledgments}

I would like to thank Dieter Brill, Andrew Chamblin, Ted Jacobson,
Stephen Hawking, Andrei Linde, Jens Niemeyer, Steve Shenker and Lenny
Susskind for interesting discussions and feedback. I am grateful to
Andrei Linde and Jens Niemeyer for their comments on a draft of this
paper. This work was supported by the German National Scholarship
Foundation and BASF.

\end{document}